\documentclass[10pt,conference]{IEEEtran}
\IEEEoverridecommandlockouts
\usepackage{cite}
\usepackage{amsmath,amssymb,amsfonts}
\usepackage{algorithmic}
\usepackage{graphicx}
\usepackage{textcomp}
\usepackage{xcolor}
\usepackage[hyphens]{url}
\usepackage{breakurl}
\usepackage[linesnumbered,ruled,vlined,noend]{algorithm2e}
\usepackage{setspace}
\usepackage{tabularx}
\usepackage{multirow}
\usepackage{tablefootnote}
\usepackage{threeparttable}
\usepackage[framemethod=tikz]{mdframed}
\usepackage{booktabs}
\usepackage{color}
\usepackage{fancyhdr}
\usepackage{tcolorbox}
\usepackage[colorlinks=false,hidelinks]{hyperref}
\usepackage{balance}
\usepackage{url}
\usepackage{tikz}
\usepackage{subfigure}
\usepackage{pifont}
\usepackage{fontawesome5}
\usepackage{array}

\def\BibTeX{{\rm B\kern-.05em{\sc i\kern-.025em b}\kern-.08em
    T\kern-.1667em\lower.7ex\hbox{E}\kern-.125emX}}

\definecolor{c1}{cmyk}{0,0.6175,0.8848,0.1490}
\definecolor{c2}{cmyk}{0.1127,0.6690,0,0.4431}
\definecolor{c3}{cmyk}{0.3081,0,0.7209,0.3255}
\definecolor{c4}{cmyk}{0.6765,0.2017,0,0.0667}
\definecolor{c5}{cmyk}{0,0.8765,0.7099,0.3647}

\definecolor{lightgrey}{rgb}{0.93,0.93,0.93}

\newtcbox{\hlprimarytab}{on line, rounded corners, box align=base, colback=c3!10,colframe=white,size=fbox,arc=3pt, before upper=\strut, top=-2pt, bottom=-4pt, left=-2pt, right=-2pt, boxrule=0pt}
\newtcbox{\hlsecondarytab}{on line, box align=base, colback=red!10,colframe=white,size=fbox,arc=3pt, before upper=\strut, top=-2pt, bottom=-4pt, left=-2pt, right=-2pt, boxrule=0pt}
\newtcbox{\hlthirdtab}{on line, rounded corners, box align=base, colback=c4!10,colframe=white,size=fbox,arc=3pt, before upper=\strut, top=-2pt, bottom=-4pt, left=-2pt, right=-2pt, boxrule=0pt}

\newcommand{\dashifted}{\raisebox{0.5\depth}{\tiny$\downarrow$}}
\newcommand{\uashifted}{\raisebox{0.5\depth}{\tiny$\uparrow$}}
\newcommand{\da}[1]{{\scriptsize\hlprimarytab{\dashifted{#1}}}}
\newcommand{\ua}[1]{{\scriptsize\hlsecondarytab{\uashifted{#1}}}}

\newcommand{\dd}[1]{{\scriptsize\hlprimarytab{#1}}}
\newcommand{\uu}[1]{{\scriptsize\hlsecondarytab{#1}}}

\begin{document}

\title{How far are AI-powered programming assistants from meeting developers' needs?}

\author{
    \IEEEauthorblockN{
        Xin Tan,
        Xiao Long,
        Xianjun Ni,
        Yinghao Zhu,
        Jing Jiang,
        Li Zhang
    }
    \IEEEauthorblockA{Beihang University, Beijing, China\\
    Email: \{xintan, longxiao, nixianjun, zhuyinghao, jiangjing, lily\}@buaa.edu.cn}
}


\maketitle

\begin{abstract}

With the emergence of In-IDE AI coding assistant tools (ACATs) like \textit{GitHub Copilot}, developers' coding habits have undergone significant changes. While some studies have examined the effectiveness of these tools, there lacks in-depth investigation into the actual assistance process they provide. To bridge this gap, we simulate real development scenarios encompassing three typical types of software development tasks and recruit 27 computer science students to investigate their behavior with three popular ACATs. Our goal is to comprehensively assess ACATs' effectiveness, explore characteristics of recommended code, identify reasons for modifications, and understand users' challenges and expectations. To facilitate the study, we develop an experimental platform that includes a data collection plugin for VSCode IDE and provides functions for screen recording, code evaluation, and automatic generation of personalized interview and survey questions. Through analysis of the collected data, we find that ACATs generally enhance task completion rates, reduce time, improve code quality, and increase self-perceived productivity. However, the improvement is influenced by both the nature of coding tasks and users' experience level. Notably, for experienced participants, the use of ACATs may even increase completion time. We observe that ``edited line completion'' is the most frequently recommended way, while ``comments completion'' and ``string completion'' have the lowest acceptance rates. We also notice that the primary reasons for modifying recommended code are disparities between output formats and requirements, flawed logic, and inconsistent code styles. In terms of challenges and expectations, optimization of service access and help documentation is also concerned by participants except for functionality and performance. Our study provides valuable insights into the effectiveness and usability of ACATs, informing further improvements in their design and implementation.

\end{abstract}

\section{Introduction}
Artificial intelligence (AI) is making significant changes to our daily lives. Especially with the prosperity of generative AI, some prominent AI assistants like \textit{ChatGPT} emerged~\cite{fui2023generative}. Their exceptional performance in natural language understanding, task assistance, and personalized responses has made them indispensable tools in various domains~\cite{babu2024revolutionizing,tian2023chatgpt,biswas2023role,chen2023gptutor}. 

The rapid advancement of AI technology has revolutionized the coding habits of developers as well. AI coding assistant tools (ACATs) have been proposed and adopted by developers in their daily development workflows~\cite{baltes2018towards,bird2022taking,pantelimon2023improving}. As reported by Microsoft CEO, by Oct. 2023, the subscription count for \textit{GitHub Copilot} has surpassed one million users, and its adoption extends to over 37,000 organizations globally~\cite{Ray2023Microsoft}. It can be seen that ACATs has become an important coding companions for developers nowadays~\cite{mcnutt2023design}. 
Unlike traditional coding assistant tools, ACATs utilize machine learning models trained on vast amounts of code (e.g., substantial repositories on GitHub). This makes them suggest more lines of code not only based on the developer's input but also considering the code context or prompts~\cite{zan2023large, yetistiren2022assessing}. Thus, ACATs are capable of providing more intelligent, context-aware, and personalized code suggestions compared to traditional tools~\cite{ross2023programmer,ponzanelli2016prompter}.

Several ACATs, such as \textit{GitHub Copilot}~\cite{GitHubCopilot_Feature}, \textit{Tabnine}~\cite{Tabnine_Feature}, and \textit{CodeGeex}~\cite{zheng2023codegeex}, have been developed and made available as IDE plug-ins for developers to download and utilize. These tools have gained significant popularity among developers, with their providers highlighting their ability to greatly enhance productivity~\cite{wermelinger2023using,corso2024assessing,li2023exploring}. For instance, \textit{GitHub Copilot} claims that developers who use it complete tasks 55\% faster than those who do not~\cite{Kalliamvakou2022quantifying}. 
However, some recent studies have failed to detect any significant disparities in task completion and code quality when employing ACATs~\cite{10.1145/3487569}. The influence of ACATs on developers is also multifaceted~\cite{perry2023users,golden2018differences}. Therefore, relying solely on metrics like task completion time or completion rate may not fully capture the significance of tools in facilitating a cohesive programming experience and developers' self-perceived productivity~\cite{10.1145/3558489.3559072, al2022readable}.
Given their widespread adoption and diverse opinions from users, developers are curious about whether these tools really as effective as their provider claims and which aspects can be improved. Thus, a critical question arises - what makes a good AI coding assistant tool?

To evaluate the effect of ACATs, researcher have proposed many metrics and benchmarks, e.g., \textit{Exact Match}~\cite{haque2022semantic}, 
\textit{CodeBLEU}~\cite{ren2020codebleu}, and \textit{Pass@k}~\cite{chen2021evaluating}. These approaches primarily focus on evaluating the correctness or similarity of the generated code~\cite{takaichi2022nlp}, without 
considering the interaction between developers and ACATs. However, the interaction is an important aspect for evaluating the performance of these tools, 
which can provide insights into how well the ACAT integrates into the developer workflow and how effectively it supports developers in their coding tasks~\cite{chopra2024exploring, sheese2024patterns}. 
Some recent studies~\cite{liang2023large,jaworski2023study, ziegler2022productivity,wang2023practitioners, sobania2022choose} have begun to focus on user experience, but most of them are based on user surveys, lacking detailed analysis of the actual assistance process. Although a few studies~\cite{10.1145/3491101.3519665,10.1145/3491102.3501870,10.1145/3487569} conduct user experiments, they still lack fine-grained analysis of the assistance process. For example, they do not consider different types of recommendation granularity and the corresponding requirements, challenges, and expectation during coding process. This knowledge can bring rich insights for the future improvement of ACATs.

In this paper, we take one step towards addressing this gap. We try to simulate real software development scenarios and conduct a controlled human study with 27 computer science students. We invite them to participate in three typical software development tasks (i.e., Algorithms and Data Structures, Management System Development, and Research Tool Development) with/ without the help of three popular ACATs (i.e., \textit{GitHub Copilot}, \textit{Tabnine}, and \textit{CodeGeeX}). To facilitate this study, we develop an experimental platform that integrates a data collection plugin for the VSCode IDE and 
supports task 
description, 
timekeeping, code evaluation and generation of personalized interview and survey questions. Through the comprehensive analysis of developers' submitted code, developer behavior data and results of interviews and surveys, we reveal the effectiveness of different ACATs in different tasks, as well as the scenarios that developers use ACATs and corresponding challenges they may encounter. 

We find that ACATs increase task completion rates, resulting in a general reduction in task completion time. However, ACATs even pose as a hindrance for experienced users. ACATs also enhance code quality and participants’ self-perceived productivity. However, the impact varies across different tasks, with ``Management System Development'' being the most affected. The characteristics of the recommended code indicate that ``edited line completion'' is the most frequently recommended way, while ``comments completion'' and ``string completion'' have the lowest acceptance rates. The study also highlights 22 factors influencing developers' evaluations and decisions of ACATs, as well as 9 challenges that users may faced, such as limitations in handling complex logic and varying preferences for code recommendation length. Lastly, participants express 23 expectations for ACATs, including optimized code suggestion ranking, improved debug accuracy, and adaptive learning of coding style. Overall, these findings contribute to the understanding and optimization of ACATs, thereby enabling them to provide improved support to developers and enrich their programming experiences. 

\section{Related Work}
We discuss 
studies on ACATs' usability and developers' interaction with them. Since this field is rapidly developing, the following is a snapshot  
as of February 2024.

Some studies compare human programmers with ACATs. 
Imai~\cite{imai2022github} 
conducts an experiment 
with 21 participants to compare the productivity and code quality of pair programming with \textit{GitHub Copilot} versus human pair programming. 
They find that while programming with \textit{Copilot} generates more lines of code, 
 the code quality is lower. Nascimento et al.~\cite{nascimento2023comparing} 
compare the performance of software engineers and \textit{ChatGPT} on non-functional evaluation metrics. 
\textit{ChatGPT} 
surpasses novice programmers in solving easy and medium-level problems, but lacks evidence to 
outperform experienced 
programmers. Dakhel et al.~\cite{dakhel2023github} compare \textit{Copilot} with human programmers on solving fundamental algorithmic problems. 
They find that humans' solutions have a higher correct ratio, 
but \textit{Copilot}'s buggy solutions are easier to fix. 
Instead of comparing programmers and ACATs in a laboratory environment, we 
conduct a user study to treat ACATs as an assistant rather than a programmer alternative, which is more closer to real-world software development scenarios.

Some studies investigate how ACATs affect developers to complete certain programming task~\cite{vaithilingam2022expectation, 10.1145/3487569}. Both studies find that the time difference of completing tasks bring by ACATs is not significant. Other recent studies~\cite{ross2023programmer,ziegler2022productivity} focus on developers' interactions with ACATs. For example, Ziegler et al.~\cite{ziegler2022productivity} find that acceptance rate is better correlated with productivity than code snippets' persistence. However, these studies lack a detailed analysis of the interaction process.

Previous studies also collect users' challenges and expectations with ACATs based on surveys or interviews. Ciniselli et al.~\cite{ciniselli2023source} survey 80 
developers on crucial characteristics of ACATs 
and construct a taxonomy of 70 ``requirements''. 
Wang et al.~\cite{wang2023practitioners} survey 599 practitioners about their expectations on code completion and compare 
with the existing research.
They find that no paper evaluates the generated code via grammatical correctness and readability, which the practitioners value most. Liang et al.~\cite{liang2023large} find that, among all challenges developers encountering, 
intellectual property and ACATs' access to their code are the top concerns. Although our study also involves interview and survey, they are conducted immediately after the simulation experiment and the corresponding questions are personalized 
based on the participants' interactions with ACATs. 
Thus, participants 
are easier to recall the specific challenges and expectations they encountered. 

Existing studies have investigated 
the effects of ACATs, usability challenges, and users' expectations, but the details of the assistant process remain unclear.
Our study validates and extends these studies by simulating real development scenarios, evaluating multiple ACATs across diverse tasks, and deeply analyzing recommendation code characteristics.   
Thus, we can gain a more comprehensive understanding of ACATs.

\vspace{-0.1cm}
\section{Study Design}
\begin{figure}[htbp]
    \centering
    \includegraphics[width=1\linewidth]{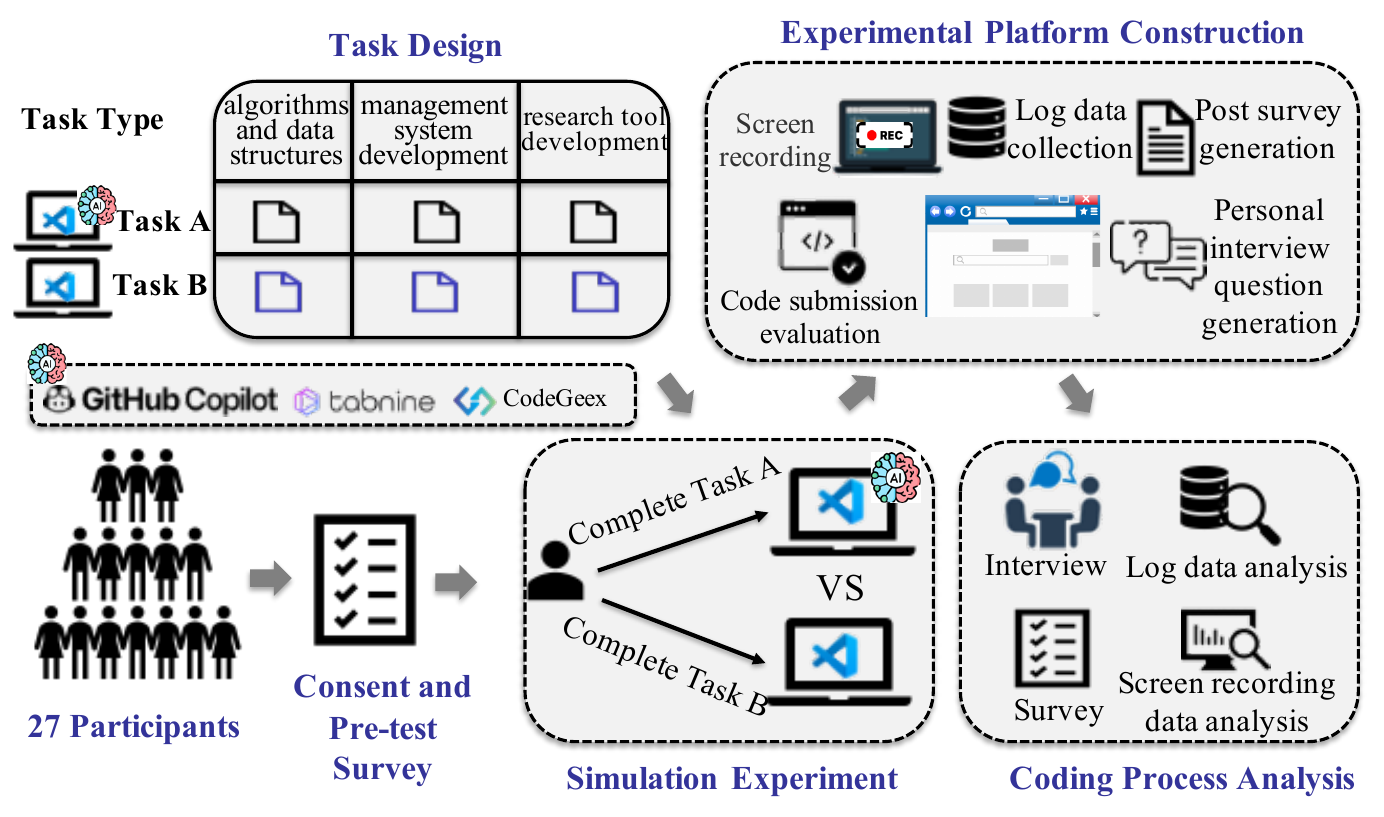}
    \vspace{-0.7cm}
    \caption{Study Overview}
    \vspace{-0.6cm}
    \label{fig:study_overview}
\end{figure}

\vspace{-0.1cm}
This study aims to assess ACAT effectiveness across scenarios and developer interactions, elucidating fine-grained user requirements and challenges. Insights gained can inform future ACAT development. To achieve this goal, we conduct the controlled human study as shown in Fig~\ref{fig:study_overview}. We design three typical types of coding tasks including two sub-tasks for each type and consider three popular ACATs. We ask participants to complete one sub-task using VSCode with an ACAT plugin and the other sub-task without an ACAT plugin. To facilitate our study, we construct an experimental platform that can collect the process log data and screen recording data, evaluate 
code submission, and generate personalized questions based on the participant's coding process. After 
finishing coding tasks, we immediately conduct an interview and survey with participants. 
After that, we integrate and synthesise the above multiple types of data 
to extract the findings. 

\vspace{-0.1cm}
\subsection{Construction of Research Questions}
\vspace{-0.1cm}
We first outline the research questions and illustrate the motivations behind each question.

\textit{\textbf{RQ1: How effective are ACATs?}} 
RQ1 investigates the multifaceted impacts of utilizing ACATs on development productivity in typical development scenarios, e.g., efficiency, quality, frequency , and self-perceived productivity. This investigation aims to provide a comprehensive understanding of how ACATs contribute to the productivity in different scenarios.


\textit{\textbf{RQ2: What are the characteristics of the code recommended by ACATs?}} RQ2 aims to examine the characteristics of code snippets recommended by ACATs. This analysis can provide insights into the current capability boundaries, strengths, and limitations of ACATs, enabling a better understanding of their effectiveness.


\textit{\textbf{RQ3: What factors influence users' evaluations and decisions of ACATs?}} RQ3 aims to explore users' perception of effective ACATs and reveal why they choose to accept, reject, or modify certain code recommendations. Insights derived from their actual needs and preferences can inform ACAT design and enhance its effectiveness.


\textit{\textbf{RQ4: What are the challenges and expectations when using ACATs?}}
RQ4 aims to explore the challenges faced by ACAT users and their expectations. 
Understanding these challenges and expectations can help in improving the design and functionality of ACATs, addressing the specific pain points and concerns of users, and ultimately enhancing their overall experience and productivity when utilizing such tools.

\subsection{Task Design}
To simulate real-world development activities and align with the scope of a user study, we design three types of Python coding tasks commonly encountered by computer practitioners in daily software development, as shown in Table~\ref{tab:task_design}. 
The detail information of these tasks can be accessed in the replication package. Each type has two sub-tasks (tasks A \& B), enabling us to conduct the controlled experiments. To ensure a relatively fair comparison, we try to maintain a similar level of difficulty and workload across the sub-tasks. 

The first type is Algorithms and Data Structures, the most fundamental skill for developers, essential in both university courses and job interviews. We choose two algorithm problems from Codeforces, a world-renowned online assessment system.~\cite{CodeForces_Feature}. Both problems have a difficulty rating of 1,100, with similar pass rates 
to ensure consistency in difficulty. Before the experiment, we conduct tests using three ACATs on both problems but no correct codes are generated. We confirm that the two problems are not in ACATs' training dataset.
To prevent direct search engine solutions, we keep the core algorithm but modify the problem context and description. 

The second type, Management System Development, represents the task of object-oriented and system-level development, which is 
often implemented by developers in the industry~\cite{brusa2018systems}. 
The last type is Research Tool Development, involving design and develop analysis scripts. We control the number of similar function points need to implement for consistent difficulty. 

To mitigate any deviation caused by the experimental sequence, we furnish participants with preparatory materials like half-completed code and tutorials for processing json files. We also provide the pre-defined functions for common elements across both tasks A\&B, e.g., string handling logic. 

Before formally starting the experiment, we conduct pilot experiment with three students to identify any unreasonable or unclear settings. After collecting their feedback and making improvements to our task descriptions and design, we further ensure consistency in task difficulty.



\begin{table}[htbp]
\centering
\scriptsize
\setlength\tabcolsep{3pt}
\renewcommand{\arraystretch}{1}
\caption{Three Types of Python Coding Tasks}
\vspace{-0.2cm}
\label{tab:task_design}
\begin{tabular}{p{1.6cm}p{3.2cm}p{3.2cm}}
\toprule
Category &
  \multicolumn{1}{c}{Task A (Treatment Group)} &
  \multicolumn{1}{c}{Task B (Control Group)} \\
\midrule
Algorithms and Data Structures (ADS) &
Lucky Number: \textit{judge if a number satisfies the given rule} &
Array Gracefulness: \textit{compute the best property of an array according to the given rule} \\
\hline
Management System Development (MSD) &
Merchant Class: \textit{implement merchant class of three relevant commands in a small product management system} 
  &
Customer Class: \textit{implement customer class of three relevant commands in a small product management system}
  \\
\hline
Research Tool Development (RTD) &
Bounty Issue Statistics: \textit{analyze three characteristics of substantial bounty issues from a json file} &
Top Contributor Statistics: \textit{analyze three characteristics of top contributors in an open-source community from a json file}\\
\bottomrule
\vspace{-0.9cm}
\end{tabular}
\end{table}

\subsection{ACATs Selection}
To understand how ACATs help developers, we select three most popular ACATs. They all offer code recommendations and auto-completions, utilizing AI and machine learning techniques to boost developers' productivity and efficiency. They also provide VSCode plugins. 
Table~\ref{tab:ACATs} outlines the basic information regarding these ACATs.

\begin{table}[htbp]
\centering
\scriptsize
\setlength\tabcolsep{3pt}
\vspace{-0.4cm}
\caption{Basic Information of ACATs}
\vspace{-0.2cm}
\label{tab:ACATs}
\begin{tabular}{cccccc}
\toprule
ACATs & Created-by & \begin{tabular}[c]{@{}c@{}}Release \\Time\end{tabular} & Pricing &
\begin{tabular}[c]{@{}c@{}}Downloads \\(--Mar. 2024)\end{tabular} & \begin{tabular}[c]{@{}c@{}}Version \\Studied\end{tabular} \\
\midrule
GitHub Copilot & \begin{tabular}[c]{@{}c@{}}GitHub and OpenAI\end{tabular} & 2021 & Paid & 14M+ &  v1.144.0\\
Tabnine & Tabnine & 2018 & Paid & 6.6M+ &  v3.51.0\\
CodeGeex & Zhipu AI & 2022 & Free & 661K+ &   v2.2.6\\
\bottomrule
\vspace{-0.7cm}
\end{tabular}%
\end{table}

\subsection{Participants Recruitment}
We recruit participants from the pool of computer science students at our university. To screen eligible applicants, we conduct a pre-test survey to collect their self-reported Python proficiency, experience with ACATs, and familiarity with each of the three task categories. The details of pre-test survey can be accessed in the replication package. We only recruit the applicants with some experience in Python (i.e., $\geq3$, [1: very inexperienced] to [5: very experienced]). Finally, 27 participants are recruited, and each individual signed a consent form prior to their participation, which grants us permission to collect log data, screen recording data, as well as conduct interview and survey. Participants receive ¥250 as compensation for their time. Our study has undergone a thorough review by the Ethics Committee of our university and has been approved.

\subsubsection{\textbf{Demographics}}
Our 27 participants include 6 females and 21 males, and 10 participants are senior undergraduate students, the other 17 are master students. They all major in computer science. Regarding their Python experience, 18 participants have 2-5 years of experience, and 9 participants have over 5 years of experience.

\subsubsection{\textbf{Task Assignment}}
To reduce the impact of participants unfamiliarity with tasks and tools on experiments, we assign tasks based on the experimental design and their reported familiarity levels with three task categories and three ACATs. 
In the experiment, we require each participant to complete task A of a certain type using one of the ACATs. To form a control group, we also ask the same participant to complete task B of the same type using only the basic functions provided by VSCode. To ensure diversity and accuracy of the data, we invite three different participants to participate in each set of experiments. For convenient data recording and organization, we use the format of ``\textit{toolName\_taskName\_index}'' to denote the participants, such as \textit{COP\_ADS\_1/2/3} for participants who use \textit{GitHub Copilot} to complete the ``Algorithms and Data Structures'' task, \textit{TAB\_MSD\_1/2/3} for those who use \textit{Tabnine} for ``Management System Development'', and \textit{CG\_RTD\_1/2/3} for those who use \textit{CodeGeeX} for ``Research Tool Development''. 

Once task assignments are finalized, we require participants to perform the tasks under designated experimental conditions and record their screens at the same time. If a participant fails to complete a sub-task within 90 minutes, we ask her/him to immediately move on the next process (e.g., uploading log data and screen recordings, then starting to complete task B; if finishing task B, moving on interview and survey). This ensures the efficiency and accuracy of our experiment.

\vspace{-0.1cm}
\subsection{Experimental Platform Development}
\vspace{-0.1cm}

To support our study, we design an experimental platform. As shown in Fig.~\ref{fig:study_overview}, this platform integrates a data collection plugin for the VSCode IDE and provides functions for task statement, code submission and evaluation, timekeeping, screen recording, log data collection, and automatic generation of personalized interview and survey questions. We briefly introduce some main functions of this platform.

\subsubsection{\textbf{Code Submission and Evaluation}}
This platform allows participants to submit their code solutions files for evaluation. The platform uses preset test cases to assess the submission. If the submission fails to pass the evaluation, targeted error prompts will be provided promptly, and participants are allowed to submit and evaluate multiple times. This mechanism helps to realistically simulate the workflow and scenarios of programmers in actual development.

\subsubsection{\textbf{Data Collection during Coding Process}} 
The platform records two types of data, i.e., log data and screen recording data.
To collect interaction log data between participants and ACATs, we develop a VSCode extension called ``\textit{ccdc-plugin}'' using Typescript and two npm packages ( \textit{Yeoman}~\cite{Yeoman_Feature} and \textit{VS Code Extension Generator}~\cite{Generator_Code_Feature}). We release it to VSCode Extension Marketplace so that each participant can download it directly in VSCode~\cite{ccdc-plugin}. When conducting the experiment, participant only needs to press a shortcut key 
to accept the recommended code, the extension will capture the recommended code accepted by the user and form a \textit{\{code, timestamp\}} pair with the timestamp of the current moment and then automatically generate a json file on the user's PC to store the pair. We capture rejected and modified code by comparing ACAT's recommendations with the final code.

In addition to log data, we also ask participants record their screens from the moment they start coding until they complete the task, while the timer is running simultaneously. The recorded video, along with the timestamp information, provides data support for the subsequent analysis of interactions between participants and ACATs, as well as the automatic generation of personalized interview and survey questions.

\subsubsection{\textbf{Automatic Generation of Personalized Interview and Survey Questions}}
\begin{figure}[htbp]
    \centering
    \includegraphics[width=0.75\linewidth]{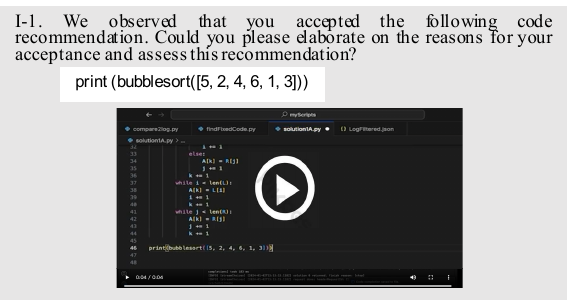}
    \vspace{-0.3cm}
    \caption{Example of Personalized Interview Question}
    \vspace{-0.6cm}
    \label{fig:interview_exp}
\end{figure}

One distinctive feature of this platform is its ability to automatically generate personalized interview and survey questions based on participants' coding processes. This is mainly achieved through the analysis of VSCode's log files, submitted code files, screen recording data, and timestamp information. Specifically, we first analyze the log files to identify the types of code snippets that participants \textit{accepted}, \textit{rejected}, or \textit{modified}, e.g., token-level, single-line level, and multi-line level. Then, we select cases from each type, capture the coding video clips corresponding to the participants at that time, and generate interview or survey questions based on our question template (refer to details in Section~\ref{sec:Interview_and_Survey_Design}). Fig.~\ref{fig:interview_exp} shows an example. The coding video clip helps participants to recall the context and mental state in which the interaction took place, leading to more realistic responses.

\vspace{-0.2cm}
\subsection{Interview and Survey Design}\label{sec:Interview_and_Survey_Design}
\begin{figure}[htbp]
    \centering
    \includegraphics[width=0.9\linewidth]{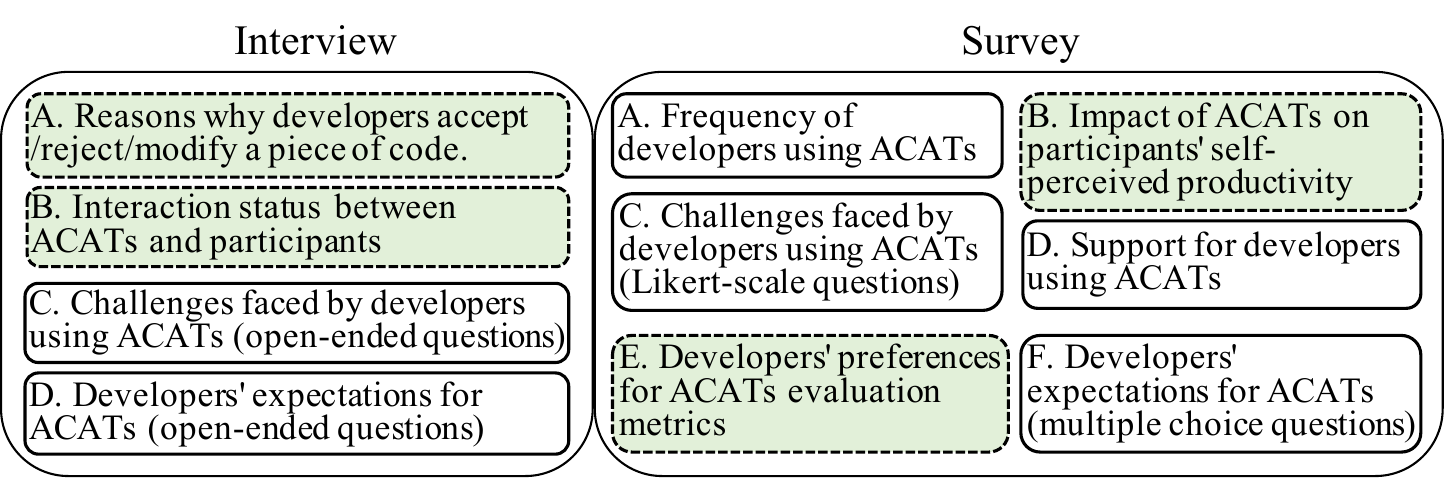}
    \vspace{-0.2cm}
    \caption{Question Dimensions in Interview and Survey (The dimensions in the dotted green frames are newly proposed.)}
    \vspace{-0.2cm}
    \label{fig:interview_and_survey_dimentions}
\end{figure}

To further 
understand how participants interact with ACATs, we conduct an after-test interview and survey with each participant. We first craft a question template, which serves as a foundation for automatic generating personalized questions tailored to each participant's coding process. 
Drawing from previous studies on human evaluation of 
ACATs~\cite{barke2023grounded,imai2022github,dakhel2023github,cheng2022would,denny2023conversing,jiang2022discovering,10.1145/3487569,puryear2022github,vaithilingam2022expectation,ziegler2022productivity}, we design a template including four established dimensions and four new dimensions, as shown in Fig.~\ref{fig:interview_and_survey_dimentions}. 
While certain dimensions overlap in both the interview and survey, the content and format of the questions differ significantly. We aim to first elicit open-ended responses through the interview, and subsequently attempt to gather more structured answers through the survey (e.g., Likert-scale question).
Then, we design specific question template 
for each dimension based on 
our study motivations. The details of interview and survey are available in the replication package. 

\vspace{-0.2cm}
\subsection{Data Analysis}
Our data comprises two types: quantitative data, e.g., task completion time, the number of accepted code recommendations, and answers to Likert-scale questions; and qualitative data, e.g., log data, screen recording data, interview data, and responses to open-ended questions.

Regarding quantitative data, we undertake descriptive statistical analysis~\cite{bajpai1992descriptive, shull2007guide}. For qualitative data, the process is more intricate. We use general best practices of qualitative analysis~\cite{merriam2019qualitative, wicks2017coding}. For screen recording data (1,909 minutes), we annotate its contents, utilizing open coding techniques~\cite{holton2007coding}. The first two authors repeatedly reviewing the videos to familiarize with the material. Then, each researcher independently assigns initial codes to identified units of analysis, reflecting their interpretations of the data. Afterward, the researchers compare their codes and engage in discussions to resolve any disparities in interpretation. Finally, based on these discussions, the researchers revise and refine their codes to reach a consensus.
For interview data, the audio recordings (225 minutes) are transcribed into textual format. Then, the first two authors apply open coding techniques to this transcribed data, following a similar process as outlined for the screen recording data.
Similar procedures are also applied to other qualitative data types (e.g., log data), ensuring a consistent and thorough analysis across all aspects of our dataset.

\section{Results}

\subsection{\textbf{RQ1: How Effective are ACATs?}}
In this question, we investigate the multifaceted impacts of utilizing ACATs on developers' productivity in typical development scenarios, including impact on tasks completion, frequency of ACAT utilization, and participants' self-perceived productivity. Each aspect contains multiple dimensions.

\subsubsection{\textbf{Impact on Task Completion}}
We evaluate ACATs' impact on task completion from three dimensions, i.e., impact on task completion rate, task completion time, and quality of submitted code. The impact of all the dimensions in this section is compared with the controlled experiments.

\textbf{Impact on task completion rate.} Table~\ref{tab:task_completion_rate} shows the task completion status of all participants. 
It reveals a notable increase in completion rates when participants perform tasks with the assistance of an ACAT. Among three task categories, the completion rate of ``Management System Development'' is most affected by the tool ($\uparrow400\%$). This observation could be attributed to the inherent nature of object-oriented class tasks, which primarily involve operations such as adding, deleting, modifying, and verifying object instances, often requiring many similar and repetitive actions. Among the three ACATs, \textit{Tabnine}'s participants achieve highest completion rate ($8/9$), which is three times higher than the controlled group. 

\begin{table}[htbp]
\centering
\scriptsize
\setlength\tabcolsep{0.9pt}
\renewcommand{\arraystretch}{1}
\caption{Number of participants who complete task and task completion rate}
\vspace{-0.2cm}
\label{tab:task_completion_rate}
\begin{tabular}{cccccccc}
\toprule
Task Category & \multicolumn{2}{c}{ADS} & \multicolumn{2}{c}{MSD} & \multicolumn{2}{c}{RTD} & Completion Rate \\ 
Sub-task & Task A & Task B & Task A & Task B & Task A & Task B & Diff\\

\midrule
GitHub Copilot & 3 & 2 & 1 & 0 & 3 & 1 & 0.44 \ua{133\%}
\\
Tabnine & 3 & 1 & 3 & 0 & 2 & 1 & \textbf{0.67} \ua{\textbf{300\%}}
  \\
CodeGeeX & 2 & 0 & 1 & 1 & 2 & 1 & 0.33 \ua{150\%}
  \\
\hline
Completion Rate & 8/9 & 3/9 & 5/9 & 1/9 & 7/9& 3/9& \\
Completion Rate Diff & \multicolumn{2}{c}{0.56 \ua{166\%}} & \multicolumn{2}{c}{\textbf{0.44} \ua{\textbf{400\%}}} & \multicolumn{2}{c}{0.44 \ua{133\%}}
  \\
\bottomrule
\vspace{-0.5cm}
\end{tabular}
\end{table}

\textbf{Impact on task completion time.} We measure the time difference between each participant completing the task with and without ACAT's help (i.e., comparing the completion time of task A against that of task B). 
If a task is not completed within 90 minutes, we record the time of completing task as 5,400 seconds. This is because we believe that the time difference bring by ACATs, rather than the time to complete specific task, better reflects the ACAT's impact on participants. The results are shown in Table~\ref{tab:task_completion_time_diff}. In most instances, ACATs decrease the average task completion time, except when \textit{CodeGeeX}'s participants completing the task ``Management System Development''. This is due to the fact that only one of the three participants in this experimental configurations complete both tasks A and B, and that he took less time to complete the task without the help of the ACAT. Among three ACATs, \textit{Tabnine} make the biggest difference ($-661s, \downarrow14\%$), with \textit{GitHub Copilot} a close second ($-641s, \downarrow13\%$). Among three task categories, the completion time of ``Management System Development'' is most affected by ACATs ($-980s, \downarrow22\%$). 

Interestingly, if we focus only on those participants who complete both tasks A and B, ACATs do not reduce the completion time. As shown in the Table~\ref{tab:task_completion_time_diff_of_participants_with_two_sub-tasks_finished}, among the seven participants who complete both tasks A and B, ACATs only reduce two participants' time (
i.e., \textit{COP\_ADS\_2} and \textit{COP\_RTD\_3}). Participant \textit{CG\_MSD\_2} 
even took 1,261 seconds ($\uparrow69\%$) longer to complete the task with ACAT assistance compared to without it. 
Thus, despite ACATs has the potential to decrease task completion time in general, for highly experienced participants in specific tasks, they may not reduce the time taken and could even pose as a hindrance.

\begin{mdframed}
[roundcorner = 4pt,linecolor = black!,linewidth = 0pt,
innerleftmargin = 3pt,innerrightmargin = 3pt,innertopmargin = 3pt, innerbottommargin = 3pt,backgroundcolor = gray!30]
\textbf{Finding 1:} We observe that ACATs increase task completion rates, resulting in a general reduction in task completion time. Among three ACATs, \textit{Tabnine} shows the best performance from these two perspectives. Nevertheless, for participants highly skilled in specific tasks, ACATs may fail to shorten the time taken and even pose as a hindrance.
\end{mdframed}

\begin{table}[htbp]
\centering
\scriptsize
\setlength\tabcolsep{2pt}
\renewcommand{\arraystretch}{1}
\caption{Time Difference of completing tasks (Measured in seconds)}
\vspace{-0.2cm}
\label{tab:task_completion_time_diff}
\begin{tabular}{cccccccc}
\toprule
Task Category & ADS & MSD & RTD & \begin{tabular}[c]{@{}c@{}}
Avg Time Diff\end{tabular} \\ 

\midrule
GitHub Copilot & -529 \da{17\%} & -457 \da{8\%} & -874 \da{16\%} & -641 \da{13\%}
\\
Tabnine & -679 \da{16\%} & -1,227 \da{23\%} & -78 \da{2\%} & \textbf{-661} \da{\textbf{14\%}}
  \\
CodeGeeX & -1,669 \da{31\%} & \textbf{420} \ua{\textbf{10\%}} & -388 \da{8\%} & -546 \da{11\%}
  \\
\hline
Avg Time Diff & -980 \da{22\%} & -421 \da{8\%} & -447 \da{9\%} & 
  \\
\bottomrule
\vspace{-0.75cm}
\end{tabular}
\end{table}

\begin{table}[htbp]
\centering
\scriptsize
\setlength\tabcolsep{2pt}
\renewcommand{\arraystretch}{1}
\caption{Time Difference of Completing Tasks of Participants with Two Sub-tasks Both Finished (Measured in seconds)}
\vspace{-0.2cm}
\label{tab:task_completion_time_diff_of_participants_with_two_sub-tasks_finished}
\begin{tabular}{cccccc}
\toprule
Participant & Time Diff & Participant & Time Diff & Participant & Time Diff\\ 

\midrule

COP\_ADS\_2 & -327 \da{11\%} & TAB\_ADS\_3 & 480 \ua{28\%} & \textbf{CG\_MSD\_2} &  \textbf{1,261} \ua{ \textbf{69\%}}
\\
COP\_RTD\_3 & -324 \da{6\%} &TAB\_RTD\_2 & 1,033 \ua{28\%}  & CG\_RTD\_3 & 241 \ua{7\%}
  \\
COP\_ADS\_3 & 226 \ua{11\%} & &
  \\
\bottomrule
\vspace{-0.6cm}
\end{tabular}
\end{table}

\textbf{Impact on code quality.}
The first two authors carefully assess the code quality of the participants' final submission. We assign a total score of three points to each task and divide its implementation into multiple steps, allocating a specific score for each step. If the code fulfill one step, it receives the all score for that step. Otherwise, the two authors score the code based on how well the code completes the command. For example, command ``\textit{addCommodity}'' in task ``Management System Development'' includes the judgement of the legitimacy of parameters such as commodity name, price, amount, user type, etc., the researchers will give their scores according to the implementation of the judgement logic of the code for these parameters. Then, we calculate the code score difference between each participant completing the task with and without ACAT's help (i.e., comparing the code score of task A against that of task B) and take an average. Table~\ref{tab:quality_score} shows score difference for each ``ACAT\_Task'' combinations. 
We observe that the average score differences are greater than 0 in all instances indicating that ACATs do improve code quality. Among three tasks, task ``Management System Development'' also has the largest score difference ($1.39, \uparrow46\%$). Among three ACATs, \textit{GitHub Copilot} brings the largest score difference ($1.61s, \uparrow54\%$).

\begin{table}[htbp]
\centering
\scriptsize
\setlength\tabcolsep{2pt}
\renewcommand{\arraystretch}{1}
\vspace{-0.6cm}
\caption{Code Quality Score Difference}
\vspace{-0.2cm}
\label{tab:quality_score}
\begin{tabular}{cccccccc}
\toprule
Task Category & ADS & MSD & RTD & \begin{tabular}[c]{@{}c@{}}
Avg Score Diff\end{tabular} \\ 

\midrule
GitHub Copilot & 1.00 \ua{50\%} & 1.84 \ua{222\%} & 2.00 \ua{200\%} & \textbf{1.61} \ua{\textbf{126\%}}
\\
Tabnine & 0.40 \ua{15\%} & 1.67 \ua{126\%} & 0.50 \ua{30\%} & 0.85 \ua{46\%}
  \\
CodeGeeX & 1.20 \ua{120\%} & 0.67 \ua{50\%} & 0.83 \ua{71\%} & 0.90 \ua{77\%}
  \\
\hline
Avg Score Diff & 0.87 \ua{47\%} & \textbf{1.39} \ua{\textbf{120\%}} & 1.11 \ua{87\%} & 
  \\
\bottomrule
\vspace{-0.5cm}
\end{tabular}
\end{table}

\begin{mdframed}
[roundcorner = 4pt,linecolor = black!,linewidth = 0pt,
innerleftmargin = 3pt,innerrightmargin = 3pt,innertopmargin = 3pt, innerbottommargin = 3pt,backgroundcolor = gray!30]
\textbf{Finding 2:} ACATs can enhance code quality, with \textit{GitHub Copilot} demonstrating the most outstanding performance. Considering multiple dimensions such as task completion rate, task completion time difference, and task score difference, task ``Management System Development'' is most affected by ACATs.
\end{mdframed}

\subsubsection{\textbf{Frequency of ACAT Utilization}}
To observe 
to what extent ACATs are involved in participants' development process, we assess the utilization frequency of 
code completion feature and other features.

\textbf{Utilization of code completion feature.} 
Utilizing our plugin ``\textit{ccdc-plugin}'', we count the number of ACAT recommendations and the number of those recommendations accepted by participants to show the extent of using code completion features. We then compute the acceptance rates, as shown in Table~\ref{tab:code_completion_acceptance}. Among three task types, ACATs achieve the highest acceptance rate ($40\%$) in task ``Management System Development'', consistent with ``Finding 2''. Among the three ACATs, \textit{CodeGeeX} exhibits the highest acceptance rate at 49\%, whereas \textit{Tabnine} demonstrates the lowest at 24\%. However, it is noteworthy that \textit{Tabnine}'s acceptance number remains comparable to the other two ACATs. 
After further analyzing contents accepted by participants, we find one possible reason 
is that \textit{Tabnine} responses so fast and even generates new recommendations with every keystroke. Due to 
coherence in programming, a significant portion of recommendations remain unnoticed by participants, thus reducing \textit{Tabnine}'s acceptance rate.
We also find that \textit{CodeGeeX} 
frequently suggests a single line of code, necessitating multiple triggers for a complete function. 
In contrast, the other two ACATs typically recommend multiple lines of code at once, resulting in a relatively higher acceptance rate for \textit{CodeGeeX}. 

\begin{table}[htbp]
\centering
\scriptsize
\setlength\tabcolsep{0.7pt}
\renewcommand{\arraystretch}{1}
\caption{Code Completion Acceptance Rate}
\vspace{-0.2cm}
\label{tab:code_completion_acceptance}
\begin{tabular}{cccccccc}
\toprule
Task Category & ADS & MSD & RTD & \begin{tabular}[c]{@{}c@{}}
Avg Acceptance\end{tabular} \\ 

\midrule
GitHub Copilot & 75/219 (34\%) & 86/194 (44\%) & 139/353 (39\%) & 300/766 (39\%)
\\
Tabnine & 67/392 (17\%) & 156/455 (34\%) & 95/493 (19\%) & \textbf{\dd{318/1,340 (24\%)}}
  \\
CodeGeeX & 117/230 (47\%) & 140/315 (44\%) & 122/230 (53\%) & \textbf{\uu{379/755 (49\%)}}
  \\
\hline
Avg Acceptance & 259/841 (31\%) & \textbf{\uu{382/964 (40\%)}} & 356/1,076 (33\%) & 
  \\
\bottomrule
\vspace{-0.3cm}
\end{tabular}
\end{table}

\textbf{Utilization of other features.} In addition to basic code-completion feature, ACATs often provide other features, e.g., using natural language to chat with ACATs~\cite{GitHub_Copilot_Chat_Feature}. We ask participants about their utilization of these features in the interviews. We observe that 12 participants out of 27 use these features, of which nine use the ``natural language to code'' feature, three use the ``debug'' feature, one participant ``inquires about API usage'', and one participant use ACATs to ``generate a test unit'' for a function he write.
\begin{mdframed}
[roundcorner = 4pt,linecolor = black!,linewidth = 0pt,
innerleftmargin = 3pt,innerrightmargin = 3pt,innertopmargin = 3pt, innerbottommargin = 3pt,backgroundcolor = gray!30]
\textbf{Finding 3:} The acceptance rate of ACATs varies with task type, peaking at 40\% for ``Management System Development''. Among three ACATs, \textit{CodeGeeX} leads with a 49\% acceptance rate, likely due to its single-line code recommendations. Conversely, \textit{Tabnine} has the lowest acceptance rate, possibly attributed to its overly fast response time. Moreover, 44\% of the participants use the non-code-completion features, particularly NL2Code.
\end{mdframed}

\subsubsection{\textbf{Participants' Self-perceived Productivity}}
\begin{figure}[htbp]
    \centering
    \includegraphics[width=1\linewidth]{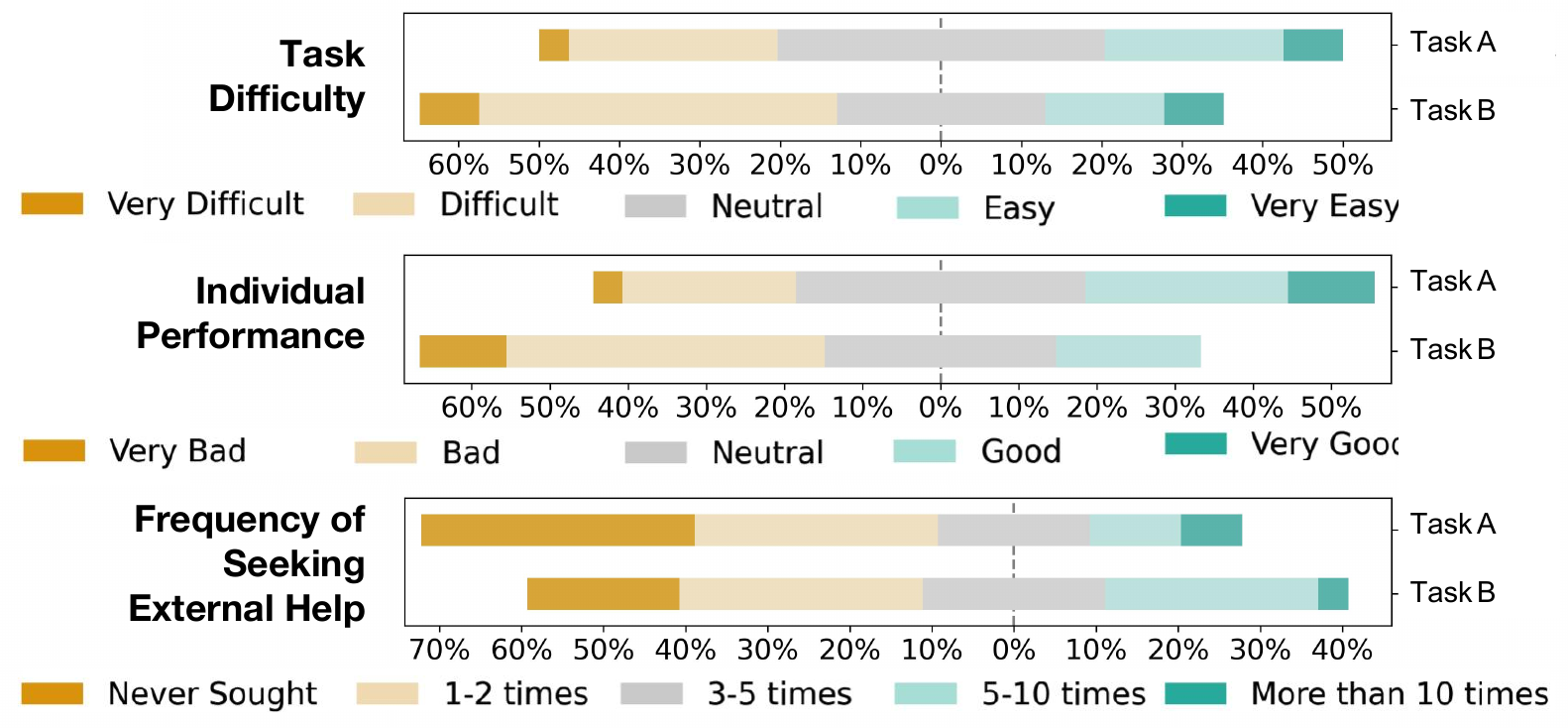}
    \vspace{-0.8cm}
    \caption{ACATs' Impact on Participants' Self-perceived Productivity}
    \vspace{-0.4cm}
    \label{fig:self-perceived_productivity}
\end{figure}

\begin{figure}[htbp]
    \centering
    \includegraphics[width=1\linewidth]{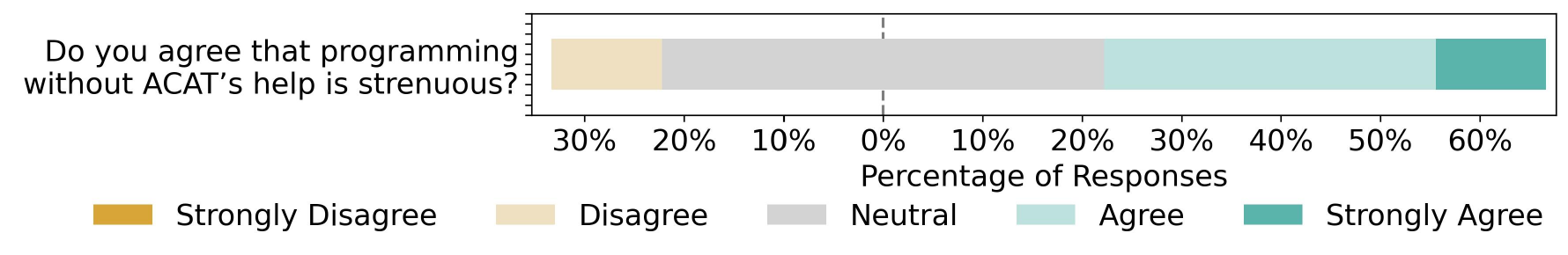}
    \vspace{-0.8cm}
    \caption{Participants' Reliance on ACAT}
    \vspace{-0.4cm}
    \label{fig:participants'_reliance_on_ACAT}
\end{figure}

ACATs' impact on participants' self-perceived productivity is often implicit and thus overlooked by many previous studies. However, it is highly correlated with the programming experience and emotional value gained by participants~\cite{ziegler2022productivity}. We explore it in the after-test survey from multiple dimensions. 
The results are shown in Fig.~\ref{fig:self-perceived_productivity}. For majority of participants, the lack of ACAT's help is accompanied by a rise in participants' perceived task difficulty, a decrease in participants' ratings of their self-performance, and a rise in times they seek external help. Fig.~\ref{fig:participants'_reliance_on_ACAT} shows participants' reliance on ACAT, indicating that 55\% of participants feel programming more strenuous without ACAT's help.

\begin{mdframed}
[roundcorner = 4pt,linecolor = black!,linewidth = 0pt,
innerleftmargin = 3pt,innerrightmargin = 3pt,innertopmargin = 3pt, innerbottommargin = 3pt,backgroundcolor = gray!30]
\textbf{Finding 4}: For the majority of participants, ACATs do reduce the perceived task difficulty, improve participants' self-performance, reduce the times of seeking external help, and generally improve the participants' programming experience. More than half of the participants ($55\%$) feel programming more strenuous without ACAT's help.
\end{mdframed}

\vspace{-0.4cm}
\subsection{\textbf{RQ2: What are the Characteristics of the Code Recommended by ACATs?}}
We examine the characteristics of the code snippets ACATs recommended from code type and code length.

\subsubsection{\textbf{Acceptance Rates of Different Types of Code}}
Building upon previous studies on the classification of ACAT-recommended code snippets~\cite{wang2023practitioners}, we extend the type of code snippets and classify all code snippets accepted or rejected by participants. These code snippets are collected using our VSCode plugin, called ``\textit{ccdc-plugin}''. The categorization process is based on two main criteria: 5 types for the recommended ways and 14 types for the code content. We then calculate the acceptance rates for specific types of code snippets aiming to provide valuable insights into the types of code recommendations that currently fall short and can be further strengthened.

\begin{figure}[htbp]
    \centering
    \includegraphics[width=1\linewidth]{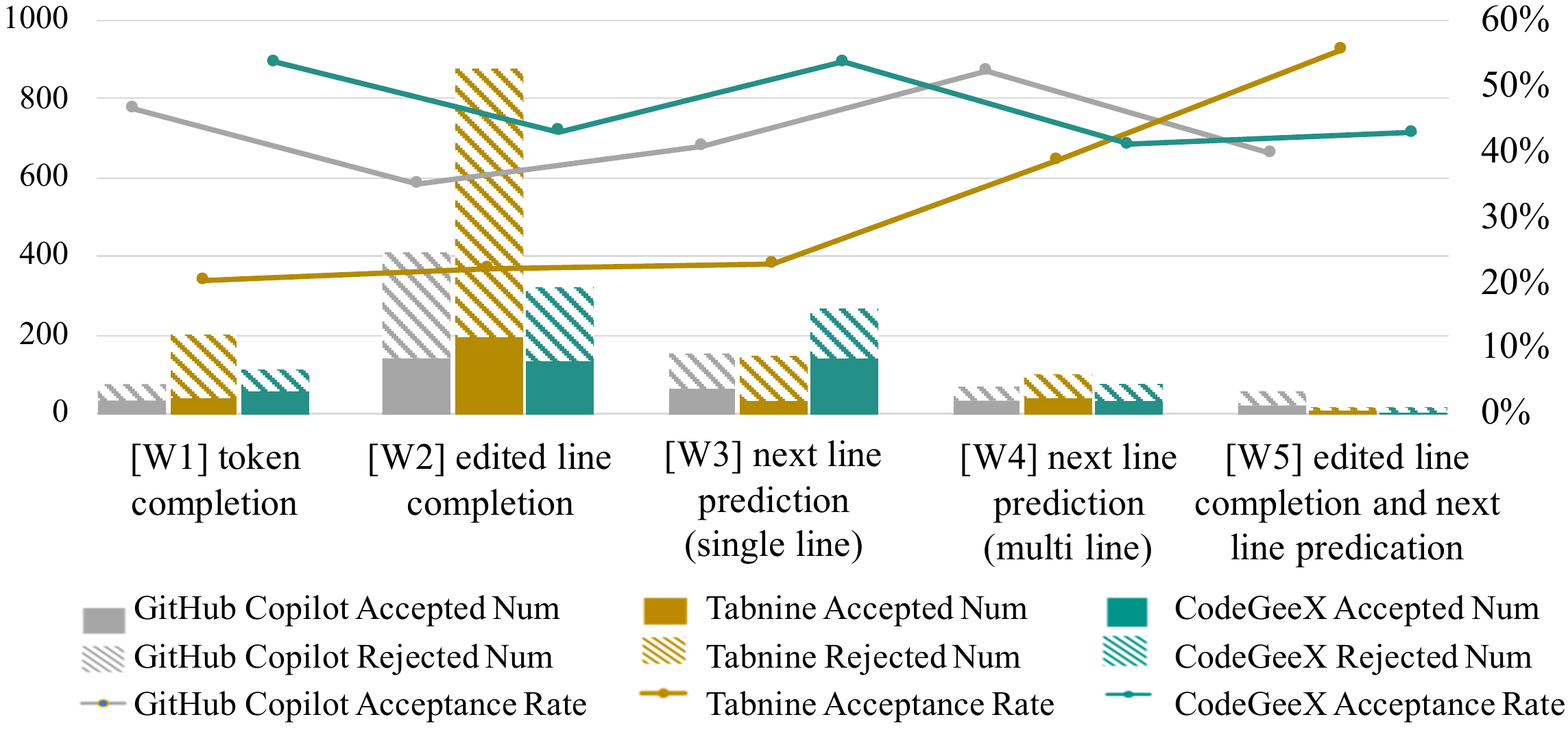}
    \vspace{-0.4cm}
    \caption{ACATs' Acceptance Rate of Different Recommended Ways}
    \vspace{-0.3cm}
    \label{fig:type_acceptance_by_methods}
\end{figure}

Fig.~\ref{fig:type_acceptance_by_methods} shows the acceptance rates of different recommended ways. We find that ``\textit{edited line completion}'' is the most frequent recommended way for all three ACATs, 
despite of its low acceptance rate. 
The two recommended ways involving multiple lines of code (i.e, W4 \&5) appear less frequently. 

\begin{figure}[htbp]
    \centering
    \includegraphics[width=1\linewidth]{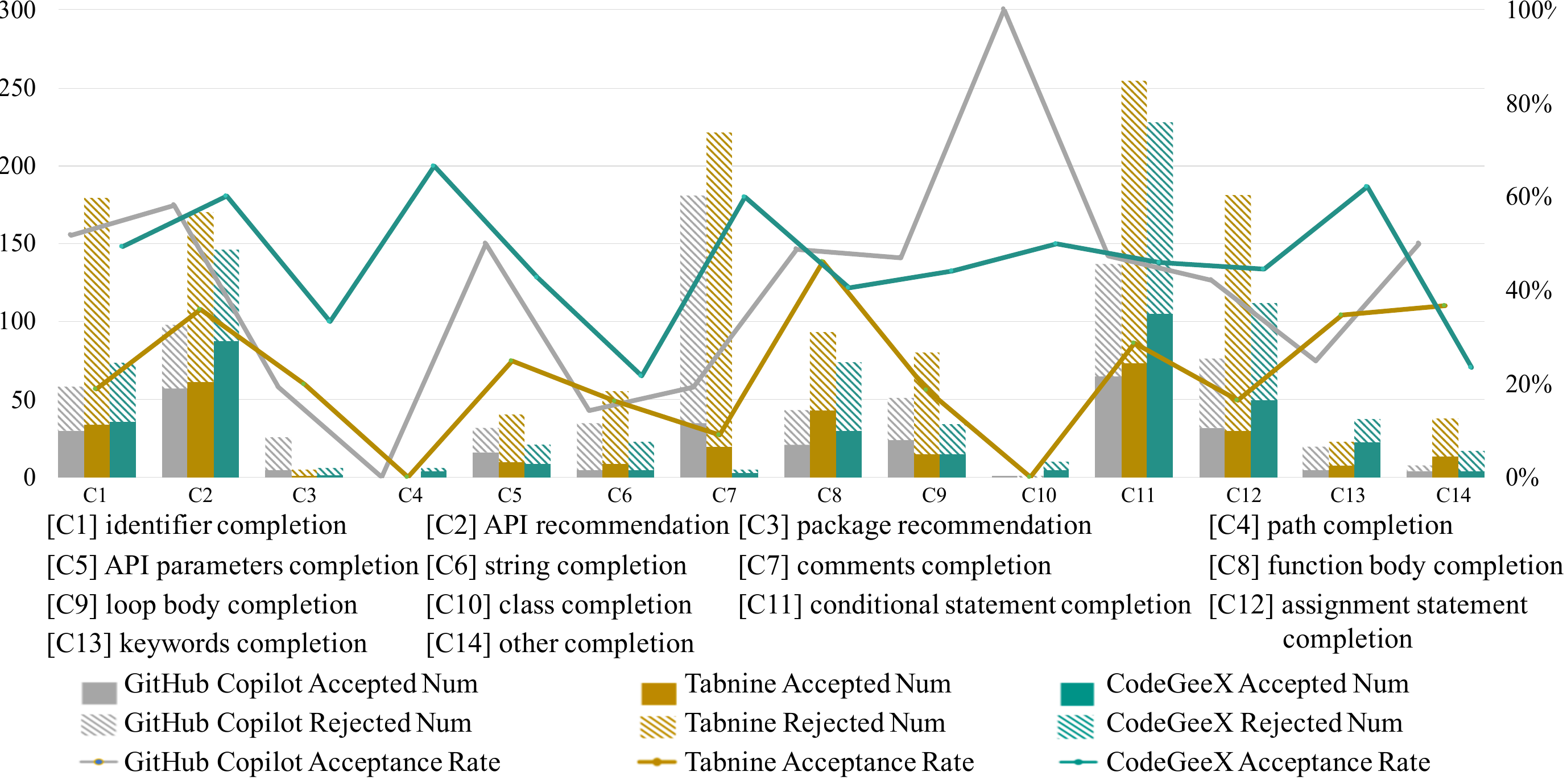}
    \vspace{-0.4cm}
    \caption{ACATs' Acceptance Rate of Different Code Content}
    \vspace{-0.3cm}
    \label{fig:type_acceptance_by_contens}
\end{figure}

Fig.~\ref{fig:type_acceptance_by_contens} shows three ACATs' acceptance rates of different code content. 
We observe that the top most frequent types are ``\textit{conditional statement completion}'', ``\textit{API recommendation}'' and``\textit{identifier completion}'', while types that are highly relevant to scenarios. For example, ``\textit{package recommendation}'', ``\textit{path completion}'', and ``\textit{class completion}'' occur less frequently in development. Excluding those types with fewer occurrences, the two types with the lowest acceptance rates are ``\textit{comments completion}'' and ``\textit{string completion}''. Each ACAT's recommended code type also has its own characteristics. For example, ``\textit{comments completion}'' appears in \textit{GitHub Copilot} and \textit{Tabnine} many times but is almost absent in \textit{CodeGeeX}. Another example is that \textit{GitHub Copilot} produces ``package recommendation'' significantly more often than the other two.

\begin{mdframed}
[roundcorner = 4pt,linecolor = black!,linewidth = 0pt,
innerleftmargin = 3pt,innerrightmargin = 3pt,innertopmargin = 3pt, innerbottommargin = 3pt,backgroundcolor = gray!30]
\textbf{Finding 5}: Regarding the recommended ways, ``edited line completion'' appears most frequently for all three ACATs, despite its low acceptance rate. Recommended ways involving multiple lines of code appear less frequently. In terms of code content, the two types with the lowest acceptance rates are ``comments completion'' and ``string completion''. 
\end{mdframed}

\subsubsection{\textbf{Average Length of Accepted and Rejected Code}}
\begin{figure}[htbp]
    \centering
    \includegraphics[width=1\linewidth]{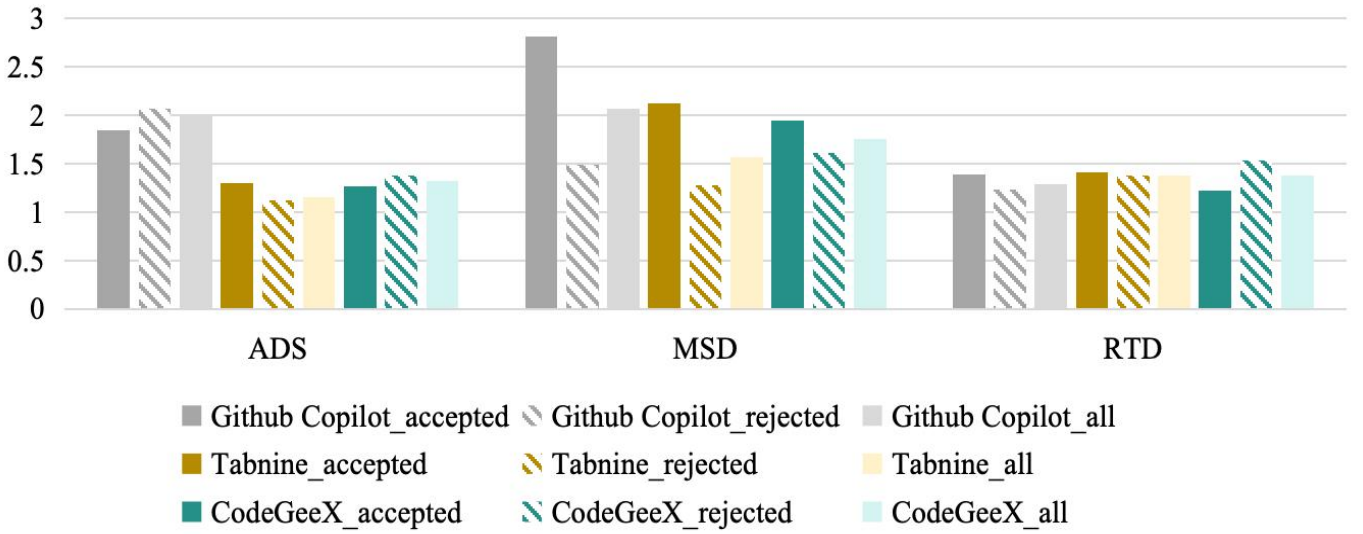}
    \vspace{-0.6cm}
    \caption{Average Length of Accepted and Rejected Code (Measured in LOC)} 
    \vspace{-0.4cm}
    \label{fig:len_of_code}
\end{figure}

We calculate the average lengths of accepted and rejected code snippets in various scenarios using the metric ``\textit{Lines of Code}''~\cite{rosenberg1997some}. The objective of this analysis is to offer valuable insights to ACAT designers regarding participants' preferences for code length. The results are depicted in Fig.~\ref{fig:len_of_code}. The average code length recommended by \textit{GitHub Copilot} is $1.97$, longer than \textit{Tabnine}'s $1.38$ and \textit{CodeGeeX}'s $1.51$. 
We also observe that the average length of code snippets accepted by participants is influenced by the type of task. Specifically, in the ``Management System Development'' task, the accepted code snippets are significantly longer compared to the other two tasks. Furthermore, the accepted code snippets are longer than the rejected ones for each ACAT, suggesting participants' preference for longer code in this task. As participants \textit{COP\_MSD\_3} said, ``\textit{I accept this because it's a long code, it even has most of the structure already written for me, which saved me a lot of time}''.

\begin{mdframed}
[roundcorner = 4pt,linecolor = black!,linewidth = 0pt,
innerleftmargin = 3pt,innerrightmargin = 3pt,innertopmargin = 3pt, innerbottommargin = 3pt,backgroundcolor = gray!30]
\textbf{Finding 6}: GitHub Copilot tend to recommend longer snippets than other ACATs. Task type influences the length of accepted code, with ``Management System Development'' tasks having longer snippets. 
\end{mdframed}

\subsection{\textbf{RQ3: What Factors Influence Users' Evaluations and Decisions of ACATs?}}
We survey participants on evaluation metrics for ACATs and interview them on their reasons for accepting, rejecting, or modifying recommended code. This combination of objective metrics and subjective opinions enhances our understanding of their views and decision-making processes towards ACATs. 

\subsubsection{\textbf{Perspectives on the Evaluation of ACATs}}
\begin{figure}[htbp]
    \centering
    \includegraphics[width=1\linewidth]{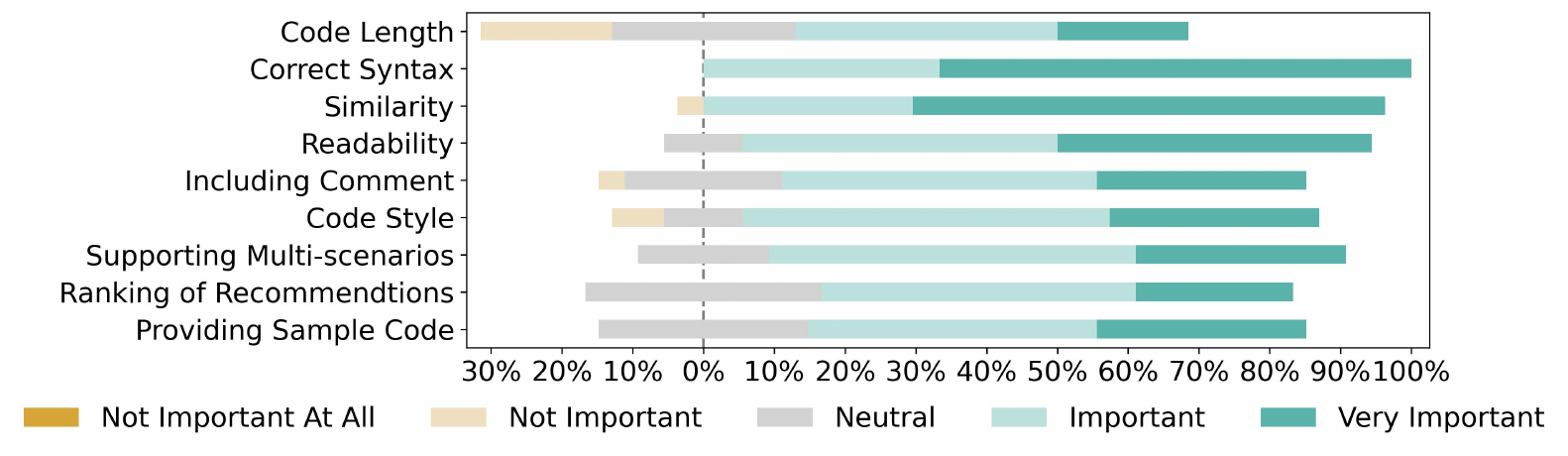}
    \vspace{-0.8cm}
    \caption{Perspectives on the Evaluation of ACATs}
    \vspace{-0.1cm}
    \label{fig:code_metric_eval}
\end{figure}

Participants offer their insights on what constitutes an effective ACAT. Fig.~\ref{fig:code_metric_eval} shows their perspectives. Among the factors expressed by the participants, ``\textit{correct syntax}'' and ``\textit{similarity to correct code}'' stand out as the most crucial factors, with over 97\% of participants emphasizing their (extreme) importance. It is because that correct syntax ensures error-free code, preventing frustrating debugging efforts for developers, while similarity to correct code ensures that the generated code aligns with the intended functionality and logic of the program. ``\textit{Readability}'' is another key concern, with 88\% of participants deeming it as (very) important. Ultimately, readable code facilitates developers' ability to quickly assess its correctness and ensures easier maintenance in the long run. Other factors such as ``\textit{consistent code style}'', ``\textit{supporting for multiple scenarios (e.g., identifiers, keywords, and multi-line completions)}'', and ``\textit{including comments}'' are all matters of concern for over 70\% of participants. Although 70\% of participants consider ``\textit{ranking of recommendations}'' (very) important, 30\% remain neutral about it, stating they disregard the rank list and often only consider the first recommendation. However, participants' opinions on ``\textit{code length}'' are diverse. 

\vspace{-0.2cm}
\begin{mdframed}
[roundcorner = 4pt,linecolor = black!,linewidth = 0pt,
innerleftmargin = 3pt,innerrightmargin = 3pt,innertopmargin = 3pt, innerbottommargin = 3pt,backgroundcolor = gray!30]
\textbf{Finding 7}: Participants offer insights on effective ACATs, highlighting factors such as ``correct syntax'' and ``similarity to correct code''. ``Readability'' and other considerations like ``support multi-scenarios'' are also important. Opinions vary on factors like ``ranking of recommendation'' (with some participants only checking the top one) and ``code length''. 
\end{mdframed}

\subsubsection{\textbf{Reasons Affecting Participants' Decisions}}

Table~\ref{tab:resaons} summarizes the reasons that influence participants' decisions to accept, reject, or modify code recommendations from ACATs. We totally identify 22 reasons. Given space constraints, we elaborate on a few of these reasons, while the majority are self-explanatory. All participants cite ``\textit{fulfilling functional requirements}'' as a reason for accepting ACATs' recommendations. This encompasses various basic code completion types, e.g., \textit{``If'' logic}, \textit{identifiers}, \textit{API recommendation}, \textit{class definitions}, and \textit{standard input and output}, indicating ACATs' general proficiency in these areas. ``\textit{Reducing keystroke numbers}'' and ``\textit{offering a framework, structure, or idea for development}'' are also frequently mentioned reasons by participants for accepting ACATs' recommendations. 

When rejecting ACATs' recommendations, participants often complain about ``\textit{generating irrelevant code, logical errors, detailed errors, or code with a different coding style}''. Another significant issue is ``\textit{missing completion due to fast keystrokes}''. One participant mentions that ``\textit{ACAT misunderstood his intention}'', as he intended to complete the code based on his comments, but the ACAT instead completed his comments. The rejection reasons underline the need for ACATs to undergo significant enhancements. These reasons indicate that ACATs should evolve beyond mere code generation to deeply comprehend developers' contexts and intentions. 

We identify eight reasons for modifying code recommendations. Notably, some of the reasons frequently mentioned overlap with those for rejecting recommendations, such as ``\textit{adjusting input/output format}'', ``\textit{enhancing logic details}'', ``\textit{correcting errors}'', and ``\textit{standardizing code style}''. Participants also highlight some intriguing reasons for modification. Three participants, for instance, mention that since the recommended code is generated based on the context code, any errors present in the context code necessitate modifications.

\begin{table}[htbp]
\centering
\scriptsize
\setlength\tabcolsep{1pt}
\renewcommand{\arraystretch}{0.9}
\caption{Reasons Affecting Participants' Decisions}
\vspace{-0.2cm}
\label{tab:resaons}
\begin{tabular}{l|l}
\toprule
Decision &
  \multicolumn{1}{c}{Reason} \\
  \midrule
Accept &
  \begin{tabular}[c]{@{}l@{}}1. Fulfilling functional requirements (27) \\ 2. Reducing the number of keystrokes required (6)\\  3. Offering a framework, structure, or idea for development (5) \\ 4. Generating simple and repetitive content efficiently (4) \\ 5. Aligning with intuitive coding practices (3)\\ 6. Providing comprehensively completed code (1) \end{tabular} \\
\hline
reject &
  \begin{tabular}[c]{@{}l@{}}1. Generating irrelevant code (4)\\  2. Missing completion due to too fast keystrokes (4)\\  3. Containing logic errors in the generated code (3)\\  4. Exhibiting detailed errors, such as output format mismatch (3)\\  5. Demonstrating varying code styles (3)\\  6. Being difficult to comprehend (1)\\ 7. Producing overly long code (1) \\ 8. Misinterpreting the developer's intention (1) \\ \end{tabular} \\
\hline
Modify &
  \begin{tabular}[c]{@{}l@{}} 1. Adjusting the input/output format to align with requirements (8) \\  2. Enhancing the logical details of the tool's recommended code (8) \\ 3. Correcting errors in the generated code (5) \\ 4. Optimizing the generated code to achieve the desired outcome (4) \\  5. Standardizing the code style (3) \\ 6. Fixing issues in the contextual code to optimize the generated code (3) \\ 7. Adjusting the complex logic of the generated code (1) \\ 8. Removing code completions accepted during debugging (1)  \end{tabular}\\
\bottomrule
\end{tabular}
 \begin{tablenotes}   
        \footnotesize         
        \item[1] Numbers in parentheses indicate \#participants mentioning this reason.
      \end{tablenotes}
\vspace{-0.3cm}
\end{table}

\begin{mdframed}
[roundcorner = 4pt,linecolor = black!,linewidth = 0pt,
innerleftmargin = 3pt,innerrightmargin = 3pt,innertopmargin = 3pt, innerbottommargin = 3pt,backgroundcolor = gray!30]
\textbf{Finding 8:} We identify 22 reasons affecting participants' decisions. They accept recommendations to fulfill functional needs, reduce keystrokes, and get development ideas. However, they reject them for irrelevant/erroneous code, and misunderstandings. Modifying reasons include error correction and contextual adjustments. 
\end{mdframed}

\subsection{\textbf{RQ4: What are the Challenges and Expectations when Using ACATs?}}
Participants express their challenges and expectations when using ACATs through interviews. Due to space restrictions, we only explain the significant or uneasy to understand ones.

\subsubsection{\textbf{Challenges}}
Table~\ref{tab:challenge} details participants' challenges with ACATs, divided into functional and non-functional categories. Functional challenges revolve around code completion, especially ACATs' ``\textit{poor performance on tasks with complex logic} (mentioned by 44\% of participants). Currently, ACATs may encounter challenges in grasping the nuances and subtleties of the code, necessitating the pursuit of developing more sophisticated algorithms and leveraging larger datasets for training. Also ranking at the top is the challenge posed by ACATs' ``\textit{poor natural language understanding capability}''.  Natural language understanding demands linguistic knowledge coupled with a profound comprehension of relevant domains or context. ACATs may fall short in this regard. Ten participants also mention the challenge of ``\textit{insufficient support for multi-type completion}''. For instance, two participants suggest that beyond basic code completion, ACATs should automatically recommend relevant packages along with API suggestions. The ``\textit{length of recommended code}'' is also a challenge. Participants express diverse opinions. Some complain that brief code recommendations may not fully meet their needs, while others find longer ones overwhelming. Hence, a pivotal question emerges: What's the ideal length of code recommendations, and how does it correlate with the context?

Non-functional challenges center around ``\textit{slow response time}'' and ``\textit{UI-related issue}'', affecting user experience. 
The first is mainly related to NL2Code feature. For the second challenge, participants mention issues including \textit{inconvenient switching methods for alternative suggestions}, \textit{indistinct auxiliary functions}, and \textit{disappearance of completion suggestions after switching interfaces}.

\begin{table}[htbp]
\centering
\scriptsize
\setlength\tabcolsep{2pt}
\renewcommand{\arraystretch}{1}
\caption{Challenges When Developers Using ACATs}
\vspace{-0.2cm}
\label{tab:challenge}
\begin{tabular}{l|l}
\toprule
Category  & \multicolumn{1}{c}{Challenge}   \\
\midrule
\multirow{5}{*}{Functional}  & 1. Poor performance on tasks with complex logic (12) \\
\multirow{5}{*}{Challenge} & 2. Poor natural language understanding capability (12) \\
& 3. Insufficient support for multi-type completion (10)  \\
& 4. Too long/short completion (3) \\
& 5. Poor performance on debug (2) \\
\hline
\multirow{4}{*}{Non-functional}
 & 1. Slow response time (10) \\
\multirow{4}{*}{Challenge} & 2. UI-related issue (8) \\
& 3. Data leakage (1) \\
& 4. Difficulty in access (1) \\
\bottomrule
\end{tabular}
 \begin{tablenotes}   
        \footnotesize         
        \item[1] Numbers in parentheses indicate \#participants mentioning this challenge.
      \end{tablenotes}
\vspace{-0.3cm}
\end{table}

\begin{mdframed}
[roundcorner = 4pt,linecolor = black!,linewidth = 0pt,
innerleftmargin = 3pt,innerrightmargin = 3pt,innertopmargin = 3pt, innerbottommargin = 3pt,backgroundcolor = gray!30]
\textbf{Finding 9}: Participants face challenges with ACATs, including poor performance on complex logic, limited natural language understanding, insufficient multi-type completion support, and varying preferences on code recommendation length. Non-functional issues like slow response time and UI concerns also arose, affecting user experience.
\end{mdframed}

\subsubsection{\textbf{Expectations}}
Participants totally express 23 expectations for ACATs, belonging to functional and non-functional categories (as shown in Table~\ref{tab:expectation})

\textbf{Functional expectation.} 
Participants express ten expectations about enhancements to current features, with over half emphasizing the desire to \textit{enhance natural language understanding and interaction capabilities}. Although it is not directly related to code completion, participants' concerns highlight the unmet need for the improvement of NL2Code functionality. The second expectation centers on ``\textit{optimized ranking of code suggestions}'', with three participants also expressing a desire to \textit{refine the UI and selection of multiple suggestions}. Sixty-seven percent of participants indicate that they review the top three code suggestions, emphasizing the importance of optimized ranking in enabling developers to swiftly identify and select the most pertinent and precise code snippets, thereby streamlining the coding process. Seven participants further hope to enhance the ``\textit{accuracy of debug functionality}''. This underscores the evolving perception of ACATs, from mere code completion tools to comprehensive assistants that can facilitate various coding aspects, ultimately elevating overall software development efficiency. Participants also mention the expectations like ``\textit{avoiding lengthy code suggestion}'' and``\textit{adaptive learning of personal coding style}''.

Participants further mention six new features that they consider necessary additions to ACATs. Five participants express the hope that ACATs can \textit{directly generate frameworks based on natural language}, particularly during the initial stages of development. This is due to the convenience and efficiency it offers, enabling developers to quickly set up the foundation of their projects using natural, intuitive language rather than spending time manually configuring frameworks. Another similar expectation is ``\textit{capability to generate project-level code}''. Some expectations are novel and interesting. For example, participants hope to \textit{enhance the multi-modal/visualization capabilities} of ACATs and to \textit{incorporate support for verbal chat}, which suggests a desire for a more interactive and intuitive coding experience. 

\textbf{Non-functional expectation.} Participants express three key non-functional expectations: accessibility, response speed, and system/UI design. Regarding accessibility, they hope that ACATs can \textit{support local connections}, minimizing the impact of network quality on their usage. Cost is another factor, with participants favoring \textit{more affordable solutions}. In terms of response speed, they desire faster code completion and NL2Code responses. Lastly, for system/UI design, participants want \textit{user support and auxiliary functions to be easily visible and accessible}.

\begin{table}[htbp]
\centering
\scriptsize
\setlength\tabcolsep{1pt}
\renewcommand{\arraystretch}{0.9}
\caption{Expectations When Developers Using ACATs}
\vspace{-0.2cm}
\label{tab:expectation}
\begin{tabular}{c|c|p{5.5cm}}
\toprule
Category   &  Sub-category & \multicolumn{1}{c}{Expectation}   \\
\midrule
\multirow{18}{*}{Functional} &
  \multirow{10}{*}{Enhancements} &
  1. Enhanced natural language understanding and interaction capabilities (15) \\
\multirow{16}{*}{Expectation} &
\multirow{8}{*}{to existing}   &
  2. Optimized ranking of code suggestions (13) \\
 & \multirow{8}{*}{features}   & 3. Improved accuracy of code autocompletion (7) \\
 &    & 4. Improved accuracy of debug functionality (7)                                   \\
 &                                   & 5. Extended length of recommended code blocks (4)                                 \\
 &                                   & 6. Clear display, selection, and configuration of multiple suggestions for code completion (3) \\
 &                                   & 7. Avoidance of lengthy code suggestions (2) \\
 &                                   & 8. Enhanced understanding of user Intent (2) \\
 &                                   & 9. Adaptive learning of personal coding style  (2) \\
 &                                   & 10. Enhanced support for Chinese language (2)\\
\cline{2-3}
 &
  \multirow{8}{*}{Adding new} &
  1. Direct framework generation from natural language requirements at initial stages (5) \\
 &  \multirow{6}{*}{features}    & 2. Capability to generate project-level code (2)                                      \\
 &                                   & 3. Programming language tutoring for beginners (2)               \\
 &                                   & 4. Addition of multi-modal/visualization capabilities (2)                            \\
 &                                   & 5. Expansion of application scenarios, including web architecture suggestions (1) \\
 &                                   & 6. Addition of verbal chat functionality (1)                     \\
 \cline{1-3}
\multirow{7}{*}{Non-functional} &
  \multirow{2}{*}{Accessibility} &
  1. Local connectivity capability (5) \\
\multirow{7}{*}{Expectation} & & 2. More affordable costs (2)  \\ \cline{2-3}
 & \multirow{1}{*}{Response} & 1. Faster completion response (6)  \\
 &  \multirow{1}{*}{speed}   & 2.  Faster NL2Code response (2) \\\cline{2-3}
 &
  \multirow{3}{*}{System/UI} &
  1. User-friendly documentation   with easy visibility (2) \\
 &  \multirow{3}{*}{Design}                                 & 2. More prominent auxiliary functions or text guidance for user convenience (1)  \\
 &   & 3. Personalized features for different user groups (1)\\   
\bottomrule  
\end{tabular}
 \begin{tablenotes}   
        \footnotesize         
        \item[1] Numbers in parentheses indicate \#participants mentioning this expectation.
      \end{tablenotes}
\vspace{-0.4cm}
\end{table}

\begin{mdframed}
[roundcorner = 4pt,linecolor = black!,linewidth = 0pt,
innerleftmargin = 3pt,innerrightmargin = 3pt,innertopmargin = 3pt, innerbottommargin = 3pt,backgroundcolor = gray!30]
\textbf{Finding 10}: Participants express 23 expectations for ACATs, including enhanced natural language understanding, optimized code suggestion ranking, improved debug accuracy, and adaptive learning of coding style. They also desire features like direct framework generation, project-level code generation, and improved multi-modal capabilities. Non-functionally, participants want local connection support, affordability, faster response times, and user-friendly design.
\end{mdframed}

\section{Threats to Validity}
\textit{Internal validity.} 
1) The tasks A\&B under each task type might affect results due to differences in difficulty and workload. For ADS-tasks, we choose two Codeforces problems with similar difficulty and pass rates. For MSD and RTD tasks, we ensure consistency in difficulty and workload by requiring the same number of similar function points to be implemented. 2) Participants' development experience may introduce bias. To mitigate this, we design a controlled experiment to investigate performance disparities in scenarios with and without ACAT assistance. 3) The sequence of completing tasks A\&B may introduce bias. To mitigate this, we provide participants with preparatory materials like half-completed code, JSON file tutorials, and pre-defined string handling functions for tasks A\&B. 4) Participants' familiarity with ACATs may also introduce bias. To address this, tasks are allocated based on their reported familiarity levels with each ACAT in the pre-test survey. 5) Participants may misinterpret the phrasing of certain survey questions. To mitigate this, we conduct a pilot survey with developers, emphasizing the clarity of the survey questions, and subsequently revise the survey based on their feedback. 6) Memory bias may threaten the internal validity, as survey questions require participants to recall their interactions with ACATs. To address this, we provide visual cues to assist participants in recalling their past experiences.

\textit{External validity.} 1) Our participants are students, who may differ from general developers in terms of experience levels, programming habits, and other aspects relevant to real-world scenarios. Future research could extend the study population for further investigation. 2) We recognize the typical challenges to reliability and generalizability when analyzing limited qualitatively data~\cite{nowell2017thematic}. We mitigate this threat by involving both authors in the coding process and including survey responses in our replication package. 3) Given the survey was conducted in January 2024, respondents' feedback reflects their ACAT interactions at that time point. Hence, some elements might not apply to subsequent tool iterations with varying functionalities and performances.

\section{Conclusion}
In this paper, we take a significant step towards understanding what makes a good ACAT. Our investigation encompasses a comprehensive analysis of ACATs' effectiveness, characteristics of their recommended code, factors influencing users' evaluations and decisions, and challenges and expectations. Through a controlled human study, we find that ACATs increase task completion rates and reduce completion time, but may hinder experienced users. They also enhance code quality and self-perceived productivity. The most recommended code completion method is ``edited line completion'', while ``comments completion'' and ``string completion'' have low acceptance rates. The study identifies 22 factors influencing users' decisions and evaluations, 9 challenges and 23 expectations for ACATs. Our study provide valuable insights for ACAT users, ACAT designers and SE researchers.

\textbf{For ACAT users}, it is recommended to actively engage with ACATs and supplementing contextual information through annotations, as this may enhance efficiency, code quality, and self-perceived productivity. However, it is important to recognize that ACATs should complement, not replace, coding skills. Users should evaluate the usefulness of recommendations and strike a balance between relying on ACATs and following their own expertise, particularly in tasks where they possess high skill. By optimizing their interaction with ACATs, users can maximize the benefits of these tools and improve their programming efficiency and effectiveness.

\textbf{For ACAT designers}, our study provides rich insights into the effectiveness of different ACATs, the characteristics of recommended code, and the challenges faced by users. This analysis helps designers understand the limitations their ACATs and provides valuable implications for future improvement. The 23 expectations expressed by users offer direct guidance for enhancing ACATs in line with user needs and expectations. By considering these findings, ACAT designers can make informed decisions and implement necessary improvements to create more effective and user-friendly tools.

\textbf{For SE researchers}, our study highlights the importance of investigating new forms of collaboration in software development, ``ACAT-Developers''. This paradigm introduces more complex factors for collaboration, e.g., tools characteristics, task characteristics, and user expertise. SE researchers could delve into these areas to gain a deeper understanding of how ACATs and human developers interact and collaborate. By exploring the nuances of this collaboration, researchers can uncover valuable insights for the effective and efficient collaborative practices in software development.

More details about our experiment design, survey and interview templates and collected data can be accessed in 
\url{https://figshare.com/s/3a475b3228be73f8a0d7}.
\balance
\bibliographystyle{IEEEtran}
\bibliography{ref}

\end{document}